%% file: paper.tex
\documentclass[11pt,preprint]{elsarticle}

\usepackage{xcolor}
\usepackage{url}

\usepackage{microtype}

\usepackage{hyperref}

\usepackage{cleveref}
\crefformat{footnote}{#2\footnotemark[#1]#3}

\newcommand{\edison}{Edisonstra{\ss}e }
\newcommand{\paul}{Paulsborner Stra{\ss}e }
\newcommand{\leibniz}{Leibnizstra{\ss}e }
\newcommand{\edi}{Edisonstra{\ss}e}
\newcommand{\pau}{Paulsborner Stra{\ss}e}
\newcommand{\lei}{Leibnizstra{\ss}e}
\makeatletter
\newcommand\footnoteref[1]{\protected@xdef\@thefnmark{\ref{#1}}\@footnotemark}
\makeatother
\journal{Journal of Pervasive and Mobile Computing}

\begin{document}

\begin{frontmatter}

\title{SimRa: Using Crowdsourcing to Identify Near Miss Hotspots in Bicycle Traffic}
\author{Ahmet-Serdar Karakaya\corref{cor1}}
\ead{ak@mcc.tu-berlin.de}
\author{Jonathan Hasenburg}
\ead{jh@mcc.tu-berlin.de}
\author{David Bermbach}
\ead{db@mcc.tu-berlin.de}

\address{TU Berlin \& Einstein Center Digital Future}
\address{Mobile Cloud Computing Research Group}
\address{Berlin, Germany}

\cortext[cor1]{Corresponding author}

\begin{abstract}
An increased modal share of bicycle traffic is a key mechanism to reduce emissions and solve traffic-related problems. However, a lack of (perceived) safety keeps people from using their bikes more frequently. To improve safety in bicycle traffic, city planners need an overview of accidents, near miss incidents, and bike routes. Such information, however, is currently not available. In this paper, we describe SimRa, a platform for collecting data on bicycle routes and near miss incidents using smartphone-based crowdsourcing. We also describe how we identify dangerous near miss hotspots based on the collected data and propose a scoring model.
\end{abstract}


\end{frontmatter}

\section{Introduction\label{sec:intro}}
\input{content/1_introduction}
\section{Related Work\label{sec:relWork}}
\input{content/6_related_work}
\section{The SimRa Platform: An Overview\label{sec:platform}}
\input{content/2a_platform}
\section{Data Acquisition\label{sec:data_acquisition}}
\input{content/2b_data_acquisition}
\section{Collected Data\label{sec:data}}
\input{content/2c_data}
\section{Data Processing and Analysis\label{sec:analysis}}
\input{content/2d_analysis}
\section{Implementation and Deployment\label{sec:implementation}}
\input{content/3_implementation}
\section{Evaluation\label{sec:eval}}
\input{content/5_evaluation}
\section{Discussion\label{sec:disc}}
\input{content/5b_discussion}
\section{Conclusion\label{sec:conclusion}}
\input{content/7_conclusion}

\section*{Acknowledgement}

We would like to thank all supporters of the project, especially TU Berlin for funding the project, Christoph Krey for implementing the iOS app version, and of course all our users.

\bibliography{bibliography}

\end{document}

%% file: content/1_introduction.tex
Major cities worldwide aim to reduce traffic emissions, traffic jams, and the city space devoted to cars.
Aside from improving public transport, the main strategy for this is to increase the modal share of bicycle traffic.
In practice, however, polls regularly show that a lack of (perceived) safety due to car-centric city planning keeps people from using their bikes more frequently.

To address this, city planners need to comprehend the ``dangerousness'' of streets at the level of individual street segments and intersections.
Such information, however, is currently not available as official accident statistics only include crashes but not near miss incidents~\cite{paper_aldred_categories}.
Yet, near miss incidents are crucial to identify dangerous segments as actual crashes are only a small subset of all dangerous situations.
Moreover, waiting for statistical significance in actual crashes would imply a large number of injured or dead cyclists, i.e., it is crucial to analyze near miss incidents instead.
Beyond using such insights to improve city planning, information on near miss incidents can also be used for dangerousness-aware routing or audio-visual warnings for cyclists approaching dangerous areas.
All such measures improve the perceived and real safety of bicycle traffic, and by that are bound to increase the traffic share of bicycles, thus, relieving traffic and environment.

Unfortunately, traditional top-down approaches are not able to collect information on near miss incidents:
Such approaches involve sending police officers to accident scenes who then file reports.
If the authorities are not notified about a near miss scene or choose not to dispatch a police officer, the near miss incident will not be included in official statistics.

Existing work on near miss incidents, e.g.,~\cite{paper_aldred_categories}, was done based on paper-based surveys with a small number of participants over a short period.
Also, a number of authors have analyzed different aspects of bicycle safety (but not near miss incidents), e.g.,~\cite{paper_wu_cyclingsafety,paper_jestico_mappingridership,paper_medury_crashcampus,paper_blanc_perceptiondemographics}.
Finally, several commercial products aim to increase the safety of individual cyclist, e.g., by automatically notifying emergency services upon a crash\footnote{https://www.tocsen.com/, https://cosmoconnected.com/en/products/cosmo-city, or https://wayguard.de/lightguard/}.
Nevertheless, all these solutions do not help to identify dangerous areas as one of the main obstacles, the lack of near miss data, has not been solved yet.

Today, the wide availability of smartphones and other mobile devices offers new ways of data collection.
In this paper, we hence propose to follow a technology-supported crowdsourcing approach to collect information on near miss incidents\footnote{In the following, we refer to near miss incidents as ``incidents''.} in bicycle traffic.
For this purpose, we developed a smartphone app that uses GPS to track routes of cyclists and built-in motion sensors to detect incidents.
Cyclists can annotate and upload their anonymized rides to our processing platform where we collect, store, and analyze the data.
Thus, we make the following contributions:
\begin{enumerate}
    \item We describe the design of the SimRa\footnote{SimRa is a German acronym that stands for safety in bicycle traffic.} platform, a crowdsourcing-based data collection and processing platform for cyclist routes and near miss incidents.
    \item We present our open source prototype which has been deployed in Berlin, Germany since March 2019, Bern, Switzerland since August 2019, and Augsburg, Bochum, Pforzheim, Stuttgart (all Germany) since mid-September 2019\footnote{Please, note that the number of regions and recorded rides is subject to change. All data presented in this paper is as of October 24, 2019}.
    \item We publish the ride data collected so far as open data.
    \item We describe a data visualization platform for such ride data and our data analysis process, including first results with insights into road safety for cyclists in Berlin.
    \end{enumerate}

This paper is structured as follows: Building on related work, we give an overview of the SimRa platform (section~\ref{sec:platform}) and describe its data acquisition process (section~\ref{sec:data_acquisition}), the collected data (section~\ref{sec:data}), and the analysis process (section~\ref{sec:analysis}).
Then, we describe our implementation and deployment (section~\ref{sec:implementation}) and evaluate SimRa based on examples (section~\ref{sec:eval}).
Finally, we discuss our approach before drawing a conclusion.

%% file: content/6_related_work.tex
Only with the broad availability of smartphones could crowdsourcing become the popular method it is today.
A prime example of such applications is collecting cycling data to improve road safety which has the potential to close gaps in official accident statistics~\cite{paper_branion-calles_comparingcrowdsourced}.

Previous work already proposes to use crowdsourcing for improving bicycle safety:
Blanc and Figliozzi~\cite{paper_blanc_orcycling,paper_blanc_perceptiondemographics} developed a smartphone application to collect information on users' cycling experience after each of their trips.
With such information, it is possible to identify streets with poor cycling experience, possibly due to safety concerns.
The app, however, does not collect data on near miss incidents or any data that could be used to identify them.
Nelson et al.~\cite{bikemaps} developed a website on which cyclist can place crash and near miss incident markers.
Similarly to them, we want to identify dangerous traffic sections.
They, however, do not collect the total ride number per section, which is necessary to determine the frequency of incidents.
Thus, it remains unclear whether a given traffic section has many reported incidents due to its dangerousness or due to the number of rides.
Moreover, their incident detection cannot be semi-automated and cyclists can report incidents at locations they have never been to.
As we will describe later, SimRa does not suffer from these shortcomings as we also record the rides and can thus consider influence factors such as the number of rides or the length of a street segment.
In the Radmesser project~\cite{ndw_2019}, 100 Berlin-based cyclists were equipped with distance sensors to identify close passes over a period of two months.
While this provides great insights, the project cannot scale and only tracks close passes.

Besides these directly related approaches, other work focuses on improving cycling safety through different methods (section~\ref{subsec:rw_goal}) or uses crowdsourcing data for a different purpose (section~\ref{subsec:rw_method}).

\subsection{Improving Cycling Safety} \label{subsec:rw_goal}

Wu et al.~\cite{paper_wu_cyclingsafety} use several open datasets such as OpenStreetMap\footnote{https://planet.openstreetmap.org/} to predict perceived cycling safety levels.
They consider sociological and environmental factors such as crime rates or the number of street lanes.
These factors have a direct and indirect influence on how cyclist perceive road safety.
Yasmin and Eluru~\cite{paper_yasmin_safetymodel} use a similar approach to predict cycling safety for specific areas.
In difference to both, we propose to use the cyclists themselves to acquire accurate, user-curated information on incidents.

Strava\footnote{https://www.strava.com/} is an application that collects crowdsourced cycling data.
Many studies rely on Strava data to analyze different cycling characteristics, e.g., ~\cite{paper_musakwa_strava,paper_jestico_mappingridership,paper_wang_spf,paper_lee_strava2,paper_hochmair_strava3}.
Furthermore, Blanc and Figliozzi~\cite{paper_blanc_orcycling,paper_blanc_perceptiondemographics} use Strava data to measure and predict cycling ridership volumes in city traffic.
The Strava data, however, is highly biased since the Strava app is mainly used for recreational activities~\cite{paper_griffin_stravalimit}.

Ferster et al~\cite{ferster_promoting_2017} describe how specifically targeting cyclists with promotion increases the amount of data submitted while targeting the general public results in a greater diversity of users in terms of age and gender.
Their results as well as results by Yang et al.~\cite{paper_yang_crowdsourcing2} could help us with recruitment of additional users.
Teschke et al.~\cite{teschke_route_2012} evaluated which of 14 road types poses the highest risk for cyclists.
Unsurprisingly, quiet streets or streets with dedicated bike infrastructure have the lowest risk of injury.

Other authors want to improve road safety in general, e.g., by automatically detecting accidents based on automated analysis of traffic flow data~\cite{ozbayoglu_real-time_2016} or through visualization of traffic data~\cite{imawan_timeline_2015}.
Using collected road safety data for route recommendation~\cite{santos_context-aware_2018} can help to reduce incidents, as well.

\subsection{Crowdsourced Data Collection} \label{subsec:rw_method}

In recent years, crowdsourcing has been used in a variety of domains, e.g., for traffic analysis~\cite{koita_crowdsourcing_2019,barka_uav_2018}, archival of analog data and digital curation~\cite{bigdata_cultural}, or location-based search~\cite{paper_chatzimilioudis_crowdsourcing}.
Furthermore, a number of research projects have developed smartphone apps for specific crowdsourcing purposes:
Le Dantec et al.~\cite{le_dantec_planning_2015} collect cycling data to make the infrastructure planning process more data-driven, e.g., to optimize the placement of bike lanes or to identify optimal synchronization schemes for traffic signals.
With the app of Aubrey et al.~\cite{aubry_crowdout:_2014}, users can report traffic offenses so that citizens improve their respect for the traffic code.
Cakmak et al.~\cite{paper_cakmak_heart} proposed a cloud-backed smartphone app to monitor heart failure patients in their daily lives.
Stevens and D'Hondt~\cite{paper_stevens_pollution} implemented a smartphone app to collect information on sound pollution.
Neither of these approaches is directly comparable to SimRa.

%% file: content/2a_platform.tex
In this section, we give a general overview of the SimRa platform which comprises all things related to the collection, storage, and analysis of crowdsourced cycling data which we will focus on in the following sections.

For data acquisition we rely on an app installed on the smartphones of participating cyclists.
This app collects data and detects incidents during rides, lets users add comments or labels, and anonymizes the data before uploading it to our servers (see section~\ref{sec:data_acquisition}).
The anonymized data
comprises information on cyclist routes, incidents, user demographics, as well as some aggregated ride statistics (see section~\ref{sec:data}).
Finally, we continuously process and analyze collected data to gain insights into dangerous street segments and intersections.
For this, we have developed one approach for interactive exploratory data analysis based on a web application and one for confirmatory data analysis~\cite{book_bermbach_benchmarking} which automatically derives a ``dangerousness'' score per street segment and intersection (see section~\ref{sec:analysis}).

%% file: content/2b_data_acquisition.tex
Our approach relies on crowdsourcing to collect necessary data; in fact, SimRa is a citizen science project.
For data acquisition, we could either rely on dedicated hardware or use commonly available hardware such as smartphones.
While dedicated hardware has certain benefits, e.g., higher measurement precision, such projects are inherently limited in scale:
\begin{figure}
    \center
    \includegraphics[width=0.5\columnwidth]{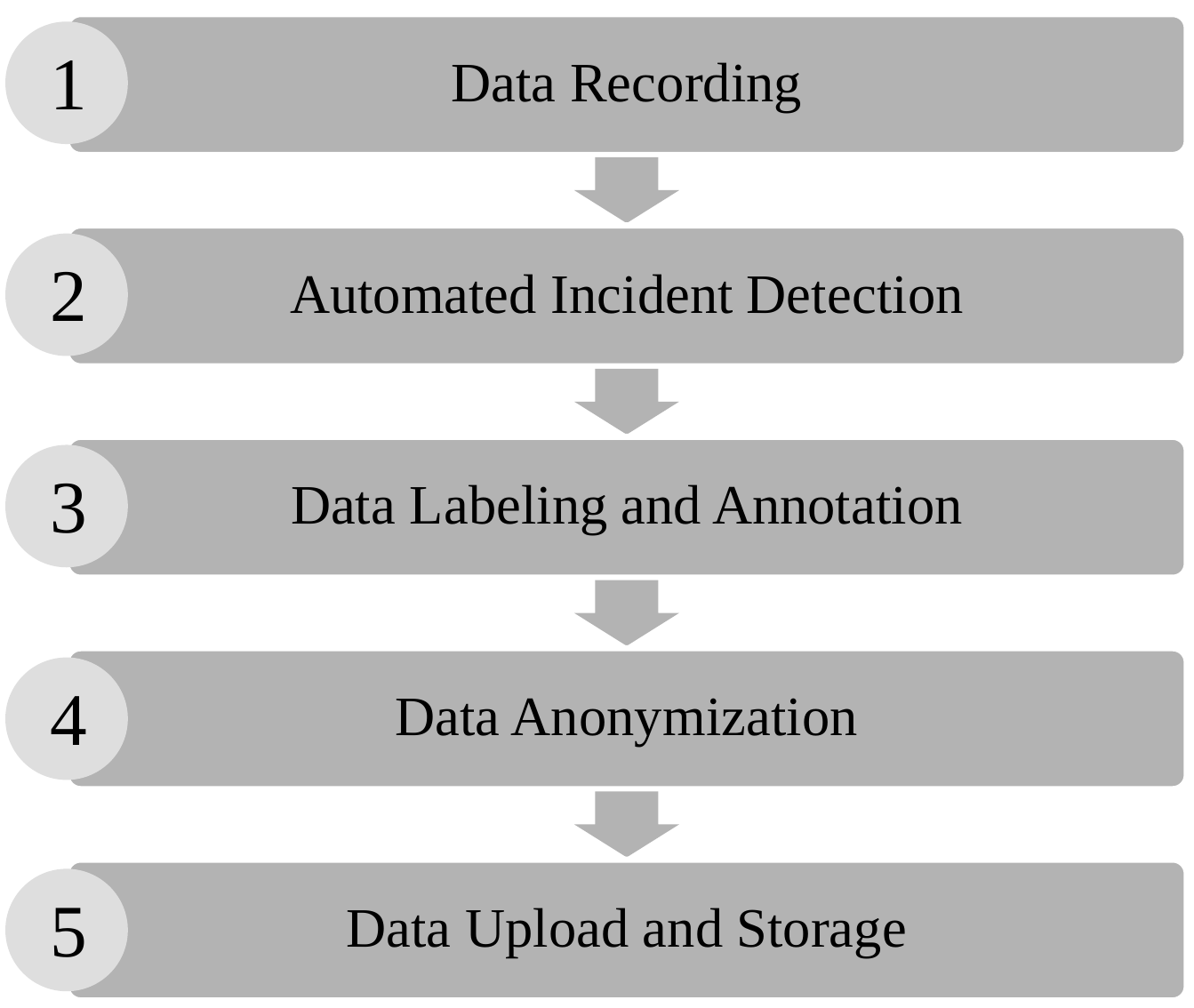}
    \caption{In the data acquisition process, collected data is manually annotated and anonymized before being uploaded to our backend servers.}
    \label{fig:data_acquisition_process}
\end{figure}
We decided for project scalability and collect data with smartphones only.
Our goal is to collect data in a way that allows us (i) to identify incident hotspots as well as the kind of incidents and (ii) to identify the routes of cyclists (in which unnecessary detours are likely to identify severe incident hotspots).

Overall, our data acquisition process has five steps and follows the structure shown in figure~\ref{fig:data_acquisition_process} (we describe the steps in the following sections in detail).
This process runs continuously and in parallel as cyclists may create data for individual rides at any time.
During a ride, we first record sensor data using the built-in sensors of a cyclist's smartphone (section~\ref{subsec:data_collection}).
Upon completion of a ride, we analyze the raw data to automatically detect incidents (section~\ref{subsec:data_incident_detection}).
Afterwards, the cyclist can enrich collected data with labels and annotations (section~\ref{subsec:data_annotation}), use a number of anonymization measures (section~\ref{subsec:data_anonymization}), and upload the data to our backend servers (section~\ref{subsec:data_upload}).

\subsection{Data Recording} \label{subsec:data_collection}
During a ride, we track three sensors at varying rates per minute.
First, we query the GPS sensor every three seconds; this returns the current location and a radius with an accuracy confidence value of 68\%.
Second, we query the smartphone's accelerometer at 50Hz.
While such a high sampling rate allows us to detect sudden peaks, this also leads to an unnecessary large data set which typically needs to be uploaded via mobile networks.
Thus, we aggregate the data based on a moving average across 30 values of which we only consider every sixth value.
This reduces the amount of data while still retaining all peaks in sensor readings.
Third, we store the device orientation based on the smartphone's gyroscope sensor every three seconds.
Each sensor measurement, together with a timestamp, is stored locally on the device.

We chose these rate settings based on initial experiments in which we identified the data collection rates and aggregation schemes as a sweet spot between system overload and information loss.

\subsection{Automated Incident Detection} \label{subsec:data_incident_detection}
After a ride, as soon as the cyclist stops the recording, we analyze the recorded data to identify incidents.
The challenge, here, is to reliably detect incidents -- initially, without any training data.

For this reason, we developed a heuristic for incident detection that relies on the assumption that incidents will often materialize as sudden acceleration spikes.
Now, that we have over 10,000 labeled rides, we started to explore alternative detection methods ranging from machine learning to signals processing.
In our heuristic, we group the acceleration time series in three-second buckets to differentiate incidents and poor road conditions (e.g., potholes result in high vertical acceleration).
In each bucket, we identify the minimum and maximum value for every dimension and calculate the difference between those two.
In a second step, we categorize the six highest difference values across all buckets as likely incidents.
This allows us to separate high acceleration values based on poor road conditions (which usually have low difference values) from incident-related peaks.

In practice, this heuristic works well for cyclists with a ``relaxed'' cycling style.
For cyclists with a more ``rapid'' style of cycling, our heuristic usually identifies either accidents, severe bumps, or traffic lights but usually not incidents.
The heuristic is also inherently limited as it cannot detect close passes and similar incidents which do not materialize as acceleration spikes.

\subsection{Data Labeling and Annotation} \label{subsec:data_annotation}

Even though we plan to improve the automated detection of incidents, some incident types cannot be automatically detected based on the sensor data alone.
For example, while being tailgated might make the cyclist ride faster, this kind of observable activity can also be related to other, non-dangerous events.
Thus, we do not think that a fully automated detection, also based on our hardware limitations, is a realistic option for SimRa -- neither now nor in the future.
Instead, we ask the cyclist to edit the pre-detected set of incidents (i.e., add false negatives and ignore false positives) and to label and annotate the correct set of incidents (see also section~\ref{sec:data} which describes the resulting data in detail).

\subsection{Data Anonymization} \label{subsec:data_anonymization}

One of our side goals in SimRa is to preserve the privacy of our users, which is mainly achieved through three mechanisms: \emph{Delayed recording} allows users to define a time and a distance threshold after which a recording will start, \emph{ride cropping} allows users to crop their ride manually to hide where they started or arrived, and \emph{per-record pseudonymization} stores demographic and ride data separately so that rides cannot be connected to individual users. Furthermore, each ride is pseudonymized separately.

\subsection{Data Upload and Storage} \label{subsec:data_upload}
Finally, and only when explicitly triggered by the cyclist, the ride data is uploaded to our backend.
For authentication, we calculate an access key based on the current timestamp and a random salt which we update with new app versions.
This is necessary to avoid automated attacks on our backend as we do not have a notion of user accounts.
So far, this has sufficed as extracting the salt from the app binary requires enough manual effort to make this infeasible for automated attacks.
Note, that we store rides and user data per region so that we can analyze (geographic) regions separately (we describe our concept of regions in section~\ref{sec:data}).

%% file: content/2c_data.tex
In this section, we describe the collected data in more detail.
For each region, we have two kinds of data sets: one for ride data (section~\ref{subsub:ride}) and one for profile data (section~\ref{subsub:profile}).

\subsection{Ride Data} \label{subsub:ride}

\begin{figure}[t]
\centering
    \includegraphics[width=0.75\columnwidth]{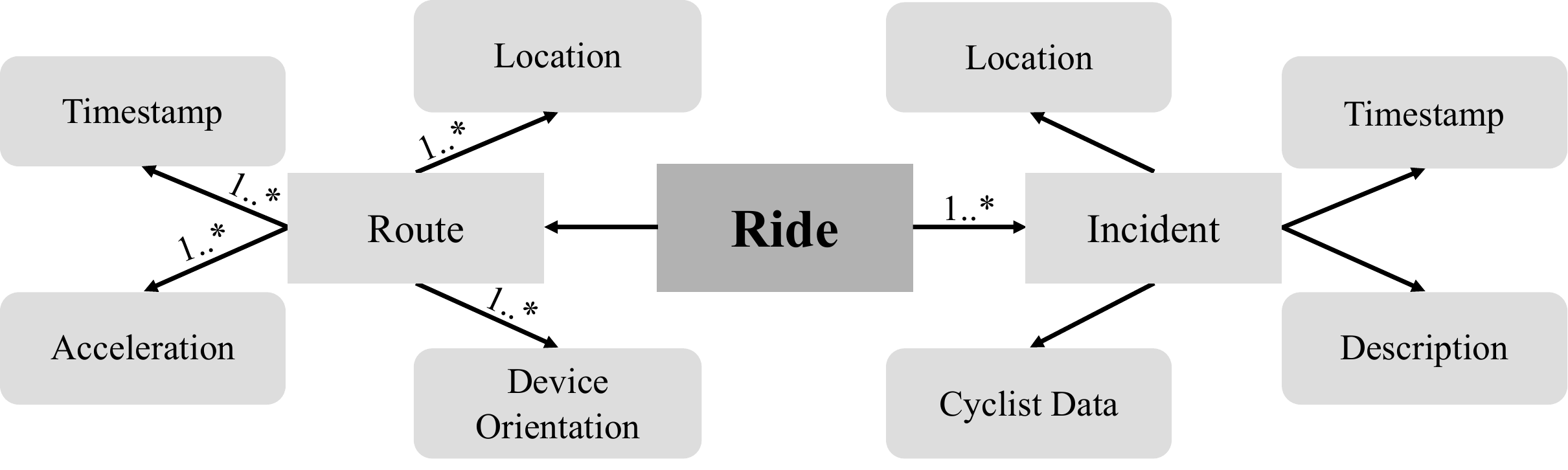}
    \caption{For each ride, we store route and incident data.}
    \label{fig:ride_data}
\end{figure}

Here, we use one file per ride; it comprises route and incident data (see figure~\ref{fig:ride_data} for an overview).
For each route, we store a sequence of geo-positions (\emph{location}) as well as \emph{acceleration} and \emph{device orientation} measurements along with the corresponding \emph{timestamp}.
Beyond this, each file also contains a list of incidents.
For each of them, we store the corresponding \emph{location} with a \emph{timestamp}, a \emph{description} of the incident, and some information on the cyclist (\emph{cyclist data}).
In the description, the cyclist can optionally describe the details of what happened, list other participants involved in the incident (e.g., taxi or pedestrian), indicate whether the incident was scary, and select an incident type.
For the incident type, we use the classification of Aldred and Goodman~\cite{paper_aldred_categories}: Close Pass, someone pulling in or out, near left/right hook, someone approaching head-on, tailgating, near-dooring, dodging an obstacle, and ``other''.
In the cyclist data, we store general information on the ride which might be useful for correlation analysis but will not identify individual cyclists.
In particular, we store the type of bike (e.g., racing bike), the location of the phone during the ride (e.g., on the handlebar), but also whether the bike had a trailer or was used to transport a child.

\subsection{Profile Data} \label{subsub:profile}

\begin{figure}[t]
\centering
    \includegraphics[width=0.75\columnwidth]{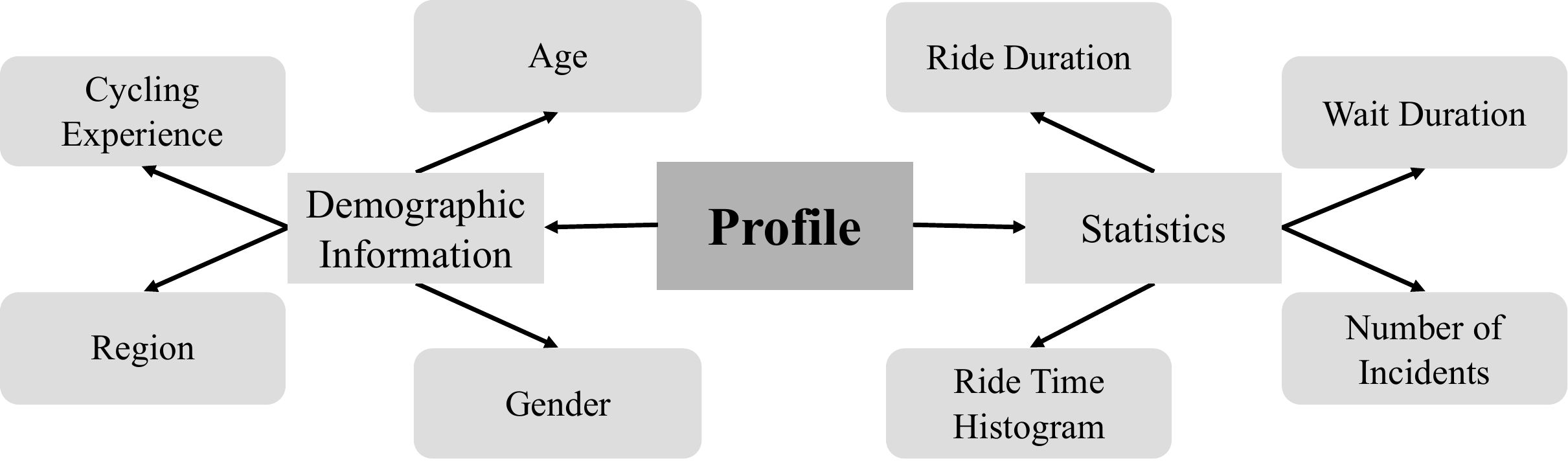}
    \caption{For each profile we store statistical and demographic data.}
    \label{fig:profile_data}
\end{figure}

In the profile data, we store demographic data and aggregated ride statistics per cyclist (see figure~\ref{fig:profile_data} for an overview).
Cyclists have one (possibly empty) profile per region.
This data is not used to identify incident hotspots but rather to support correlation analysis with demographic data that we decided not to store as part of the rides for privacy reasons.
The demographics contain the cyclist's \emph{age}, \emph{gender}, the number of years that they have cycled (\emph{cycling experience}), and the \emph{region}.
In the aggregated statics, we store the total duration of recorded rides (\emph{ride duration}), the total time spent stationary, e.g., waiting at traffic lights (\emph{wait duration}), the total \emph{number of incidents} recorded by the respective cyclist, and the \emph{ride time histogram} which shows the distribution of rides across the hours of day.

Based on this data, we could, e.g., answer the question of whether women have more incidents per km than men and whether this varies across regions.

%% file: content/2d_analysis.tex
The SimRa platform offers two options for data analysis: exploratory and confirmatory data analysis~\cite{book_bermbach_benchmarking}.
While the first is used to identify new and possibly unexpected insights in an interactive way, the latter is used to calculate predefined metrics in a fully automated way.
For exploratory data analysis, we have designed a web application which allows users to interactively explore the data set (section~\ref{subsec:exploratory}).
For confirmatory data analysis, we describe an automated analysis and scoring approach in section~\ref{subsec:confirmatory}.

\subsection{Exploratory Data Analysis} \label{subsec:exploratory}
For exploratory data analysis, we have designed a web application that allows users to interactively explore the collected data based on visual analysis.
The web application plots rides as lines and incidents as markers layered on a map.
A key feature is the set of filters through which users can filter rides and incidents based on time or ride data properties as described in section~\ref{subsub:ride}.
Based on the web application, users can identify incident hotspots by first looking for incident clusters and then comparing them to the number of rides on a particular street segment or intersection.

\subsection{Confirmatory Data Analysis} \label{subsec:confirmatory}
Our main goal is to identify incident hotspots, i.e., the most dangerous street segments, so our current approach focuses on achieving this goal.
It is, of course, possible to add further analysis goals (e.g., road surface quality).

\begin{figure}
    \center
    \includegraphics[width=0.65\columnwidth]{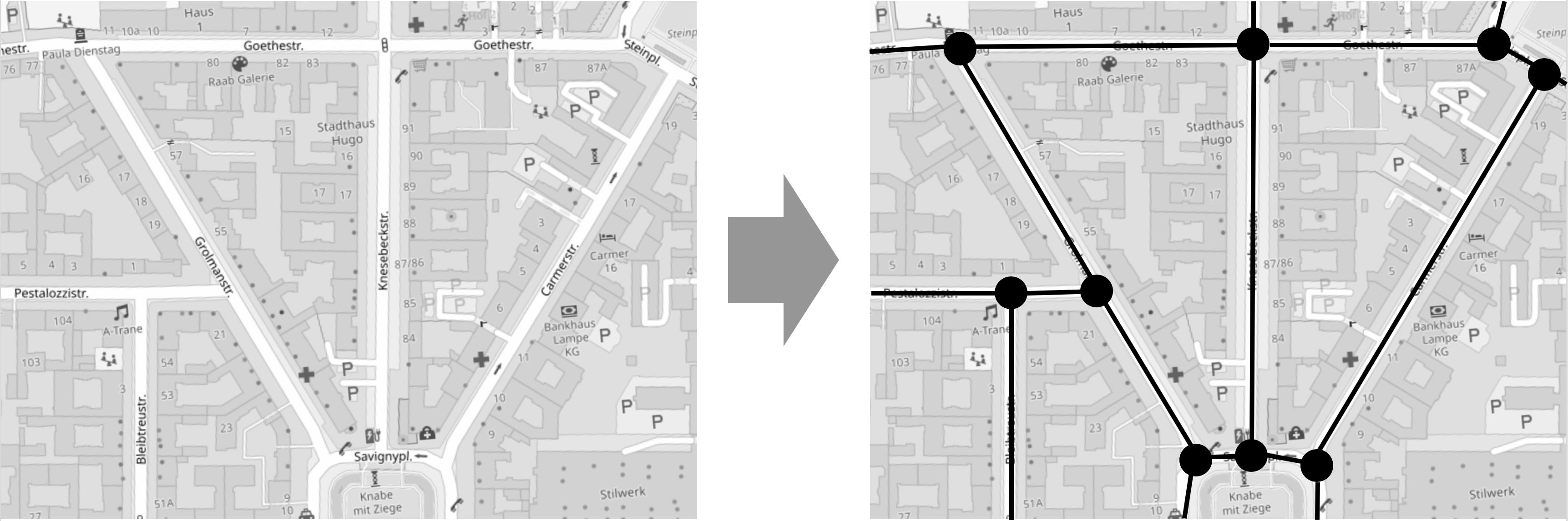}
		\caption{Example: Map and graph representation of an area in Berlin\protect\footnotemark}
    \label{fig:map_to_graph}
\end{figure}

\textbf{Data Model:} To identify incident hotspots, we first translate the collected data into a form where incidents can be mapped to intersections or street segments.
For this purpose, we have designed a simple graph model that abstracts a map: nodes represent intersections\footnote{For our purposes, an intersection is a place more than two streets meet and a street is any public way which may legally be used by cyclists.}, and edges represent street segments in between intersections.
Figure~\ref{fig:map_to_graph} shows an example of this.

\footnotetext{Map data copyrighted by OpenStreetMap and available from \url{openstreetmap.org}}
Beyond this, our graph model is in fact a weighted graph so that we can enrich it with ride and incident data.
Each node has three weights.
The first (in the following $r$) describes the number of rides which we have recorded for the respective intersection.
The second and third are vectors of length eight (there are eight incident types) and describe the number of scary ($s$) and non-scary ($n$) incidents which we recorded for the respective intersection.

Edges also use the same weights as nodes.
In contrast to nodes, however, the vectors $s$ and $n$ are eight-by-two matrices as we also need to consider the ride direction during the incident.
For the same reason, the number of rides $r$ is a vector of length two.
Finally, edges also store the length $l$ of the respective street segment as some street segments, especially in downtown areas, might be very short while others may span several kilometers.

\textbf{Scoring:} Based on this data model, we can calculate a score that describes how dangerous a particular street segment or an intersection is.
Intuitively, more incidents means higher danger score.
This, however, is not necessarily true as dangerousness is essentially the ratio of incidents to rides.
We, hence, need to consider the number of rides in our scoring function.
Furthermore, we believe that scary incidents should have more impact on the score than non-scary ones.
Hence, we propose the following formula to calculate the dangerousness score for street segments and intersections:
\begin{equation} \label{eq:score} score = \frac{1}{r} \cdot (\alpha \cdot s + n)\end{equation}

In that formula, $\alpha$ is the severity factor which affects how much more weight we assign to scary incidents than to non-scary ones.
A survey among cyclists in Berlin indicates that $4.4$ might be a good value for $\alpha$.

For street segments, $score$ is a matrix with the incident categories as rows and street direction as columns; for intersections, $score$ is a vector.
Both will usually be further aggregated based on the respective needs:
First, when the direction information of street segments is not relevant, the $score$ can further be simplified by summing up both columns pairwise.
This makes the comparison of streets and intersections easier since intersections do not have a direction in our model.
Similarly, the incident categories, i.e., the respective row entries, can be summed up so that the result is one score value for intersections and one score value per direction for street segments.
Of course, both aggregation methods can and will be combined in practice.

For improved comparability of street segments, we also need to account for the length differences, especially when the length of the shortest and the longest street segment is very different.
For this purposes, we propose a length-adjusted score for street segments:
\begin{equation} \label{eq:score2} score = \frac{1}{r\cdot l} \cdot (\alpha \cdot s + n)\end{equation}

Length-adjusted scores can be aggregated in the same way as the standard scores.
Although they are useful for better comparison of street segments, they further complicate the comparison of street segments and intersections.

Based on these scores, we can identify incident hotspots by using either a top-k approach or by manually defining a threshold score.
We suggest to only accept nodes or edges as incident hotspots if the respective number of rides exceeds a minimum, e.g., depending on the size of the data set.

\textbf{Model Population:}
For automated data analysis, the model above needs to be populated with data.
As the SimRa data set specifies incidents and rides based on a sequence of GPS locations and not based on street segments and intersections, we first need a mapping from coordinates to our model.
For this, we propose to enhance the edges and nodes in our model with a polygon each describing the geo-shape of the respective intersection or street segment.
Such data can be obtained manually for small areas or automatically from digital cadastres and public sources such as OpenStreetMap.

In a second step, the SimRa data set can then be mapped onto the model by checking for each location contained in a ride file in which polygon it is contained.
While we discard GPS values with low accuracy in the smartphone app, it may still be necessary to smooth the GPS trace.
In practice, this is usually done either based on Kalman filters or the least squares method.
Especially the latter approach is likely to be a good fit as the set of correct values, i.e., streets and intersections, is limited.

After smoothing and matching, we can easily determine how many rides are going through an intersection or street segment and can also map incidents to the corresponding street segment or intersection.

%% file: content/3_implementation.tex
\begin{figure}[t]
\centering
    \includegraphics[width=0.75\columnwidth]{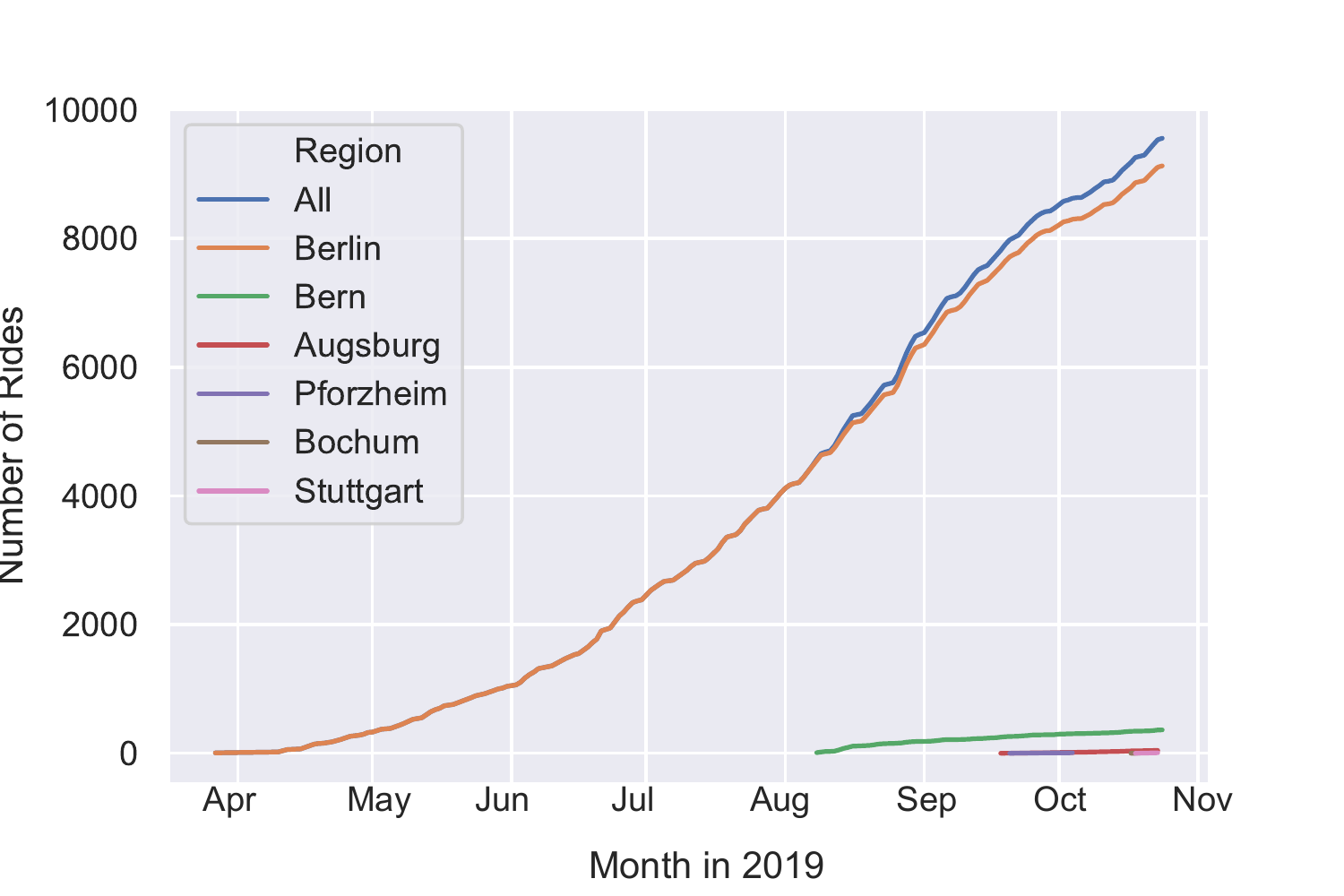}
    \caption{The total number of recorded rides is steadily growing; Berlin has the most rides as it is a major city and also the city where we initially launched the app.}
    \label{fig:rides_per_day}
\end{figure}
The SimRa platform comprises several software systems and prototypes:
We have implemented the data acquisition process from section~\ref{sec:data_acquisition} in two smartphone apps.
The Android and Lineage OS app has been implemented in Java using OpenStreetMap;
The iOS app has been implemented in Objective-C using standard platform libraries.
Screenshots of both the iOS and Android versions of the app can be found online\footnote{https://github.com/simra-project\label{note1}}.
Our backend service is a REST-based storage server implemented in Java.
Both apps and the backend service are available as open source\cref{note1}.
The processing and analysis parts from section~\ref{sec:analysis} are currently in a prototype state.
We will make them available as open source as soon as they reach a stable state.

\begin{table}[t]
\caption{Distribution of incident types across all rides as of Nov 1, 2019.}
\centering
\begin{tabular}{l|l|l}
\small Incident Type & \small Scary Incidents & \small Non-scary Incidents\\
\hline
\small Close Pass 	& \small 402 		& \small 1176 \\
\small Someone pulling in or out	& \small 71 	& \small 333 \\
\small  Near left or right hook			& \small 195 	& \small 546 \\
\small   Someone approaching head on			& \small 78 	& \small 525 \\
\small   Tailgating			& \small 42 	& \small 128 \\
\small   Near-Dooring			& \small 29 	& \small 70 \\
\small   Dodging an obstacle			& \small 110 	& \small 1519 \\
\small  Other			& \small 227 	& \small 1722 \\
\end{tabular}
\label{tab:incident_distrib}
\end{table}

For the deployment, we started recruiting citizen scientists in Berlin in September 2018.
Together with them, we defined requirements and necessary features for the app.
Using first prototypes, we gradually morphed into a beta testing phase around February 2019.
In mid-March 2019, we released the Android app in the Google Play Store\footnote{https://play.google.com/store/apps/details?id=de.tuberlin.mcc.simra.app}, deployed our backend on three TU Berlin servers, and opened the ``Berlin/Potsdam'' region.
In the last week of May 2019, we also released the iOS app in the App Store\footnote{https://apps.apple.com/gb/app/simra/id1459516968}.
Afterwards, initiatives from several cities expressed interest in using the SimRa platform.
As a result of this, we started the ``Bern, Switzerland'' region in early August 2019, as well as four more German regions in mid-September 2019 (``Augsburg'' and ``Pforzheim'') and mid-October 2019 (``Bochum'' and ``Stuttgart'').

As of today (Nov 1, 2019), we have a steadily growing number of recorded rides and -- since the curve is, at least in the summer months, slightly super-linear apparently also of users.
Figure~\ref{fig:rides_per_day} shows the total number of rides recorded with the SimRa platform since the release of the Android app
and table~\ref{tab:incident_distrib} shows the total number of incidents recorded until October 24 2019.
We continuously publish all recorded raw data\cref{note1}.

%% file: content/5_evaluation.tex
\begin{table}[t]
\caption{Base data for all street segments with \leibniz having the most rides, followed by \paul and \edison.}
\centering
\begin{tabular}{l|l|l}
\small Street Segment & \small Number of Rides & \small Length of Street Segment\\
\hline
\small \edison 	& \small 79 		& \small 230m \\
\small \leibniz	& \small 194 	& \small 628m \\
\small \paul			& \small 184 	& \small 600m \\
\end{tabular}
\label{tab:base_data}
\end{table}
As our approach is hard to evaluate in its entirety, we evaluate it based on a use case.
For this purpose, we picked three street segments in Berlin; each has a significant number of rides.
Specifically, we used our prototype for exploratory data analysis to pick three street segments in Berlin:
\emph{\edi} which appears to be an incident hotspot, \emph{\lei} which appears to be relatively safe based on the number of incidents, and \emph{\pau} which has an average number of incidents.

For each of the street segments, we calculate the incident score as described in section~\ref{subsec:exploratory} and discuss the findings from our data set based on a photo of the street segment from Google Streetview.
Specifically, we use this evaluation to also discuss the limitations of our approach.
Note, that we disregard the direction of the ride due to the (still) limited number of rides.

For our analysis, we deviated slightly from our model in that we defined our street segments based on the closest two intersections with a major street; all analysis is based on data as of Oct 24, 2019.
Table~\ref{tab:base_data} gives an overview of the resulting lengths and number of rides.

Based on these settings, we retrieved the respective number of incidents from the ride files and calculated the scores.
Tables~\ref{tab:edison}~-~\ref{tab:paul} show the resulting data, excluding incident types without occurrence.
With length-adjusted scores of about 57, 2.5, and 3.2, \edison appears by far to be the most dangerous of these streets while \paul is more dangerous than \leibniz (intuitively, these scores are the average number of incidents per meter per ride in that street segment represented as $10^{-4}$).
For each of these streets, we analyzed the results based on Google Streetview and also visited two of the three street segments.

\begin{table}[t]
\caption{Observed number of incidents and calculated score in \edison for $\alpha=4.4$}
\centering
\begin{tabular}{l|r|r|r}
\small Incident Type                                         & \multicolumn{2}{c|}{\small No. of Incidents} & \small Score \\
                                               & \small Scary                         & \small Non-Scary                 &  \small [$10^{-2}$]\\
\hline
\small Dodging an Obstacle							 & \small 18                & \small 25               & \small 131.90     \\
\hline
\small Total                                                         & \small 18                & \small 25               & \small 131.90    \\
\small Length-Adjusted Score    [$10^{-4}$]&                                      &                                    & \small 57.35 \\
\end{tabular}
\label{tab:edison}
\end{table}

\begin{table}[t]
\caption{Observed number of incidents and calculated score in \leibniz for $\alpha=4.4$}
\centering
\begin{tabular}{l|r|r|r}
\small Incident Type                                         & \multicolumn{2}{c|}{\small No. of Incidents} & \small Score \\
                                               & \small Scary                         & \small Non-Scary                 &  \small [$10^{-2}$]\\
\hline
\small Close Pass                       &                   & \small 1                & \small 0.51    \\
\small Someone Pulling In or Out        & \small 1                  &                  & \small 2.27    \\
\small Near Left or Right Hook          & \small 1                 & \small 1                & \small 2.78  \\
\small Tailgating                       & \small 1                 &                  & \small 2.27    \\
\small Near-Dooring                     &                   & \small 1                  & \small 0.51     \\
\small Dodging an Obstacle							 & \small 1                  & \small 3                & \small 3.81     \\
\small Other                            & \small 1                 & \small 2                & \small 3.30    \\
\hline
\small Total														 &\small  5                 &\small  8                & \small 15.46    \\
\small Length-Adjusted Score	[$10^{-4}$]&  								 &									&\small  2.46 \\
\end{tabular}
\label{tab:leibniz}
\end{table}

\begin{table}[t]
\caption{Observed number of incidents and calculated score in \paul for $\alpha=4.4$}
\centering
\begin{tabular}{l|r|r|r}
\small Incident Type                                         & \multicolumn{2}{c|}{\small No. of Incidents} & \small Score \\
                                               & \small Scary                         & \small Non-Scary                 & \small  [$10^{-2}$]\\
\hline
\small Close Pass                       & \small 4                 & \small 12                & \small 16.09    \\
\small Someone Approaching Head On      &                   & \small 3                & \small 1.63    \\
\small Dodging an Obstacle							 &                   &\small  1                  & \small 0.54     \\
\small Other                            &                   & \small 2                &\small  1.08    \\
\hline
\small Total                                                         & \small 4                   & \small 18                &\small  19.34    \\
\small Length-Adjusted Score    [$10^{-4}$]&                                      &                                    & \small 3.22 \\
\end{tabular}
\label{tab:paul}
\end{table}

\begin{figure}[t]
	\center
	\includegraphics[width=0.9\columnwidth]{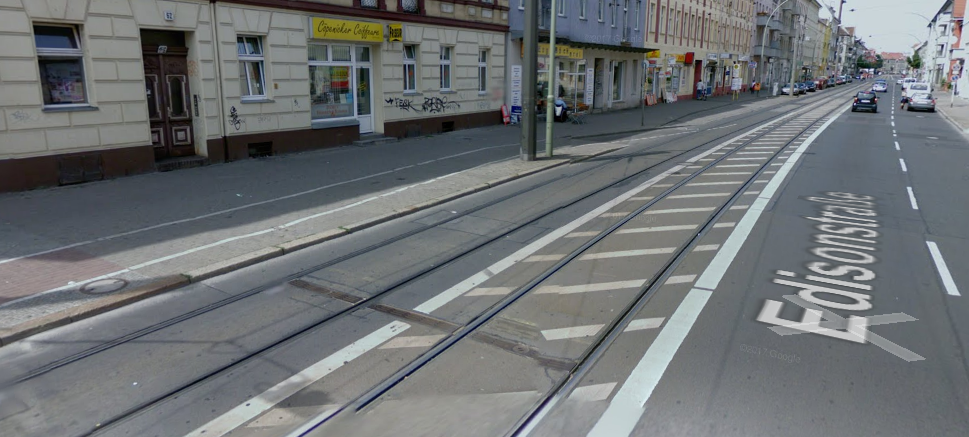}
	\caption{On \edi, illegally parked cars force cyclists to merge into high-speed car traffic on top of tram tracks (screenshot from Google Streetview).}
	\label{fig:edison}
\end{figure}

\textbf{\edi:}
Here, the main problem appears to be dodging obstacles.
When taking a look at the incident descriptions, it becomes clear that cars are frequently blocking the bike lane or are parking in second row.
When cyclists traveling towards the viewer (in the left part of figure~\ref{fig:edison}) are blocked by an illegally parked car, they are not only confronted with other cars usually driving much faster.
They also need to avoid the tracks of the tram line which can easily trip a cyclist.
Based on this, we conclude that this street segment is an incident hotspot and a very dangerous place for cyclists.

\begin{figure}[t]
	\center
	\includegraphics[width=0.9\columnwidth]{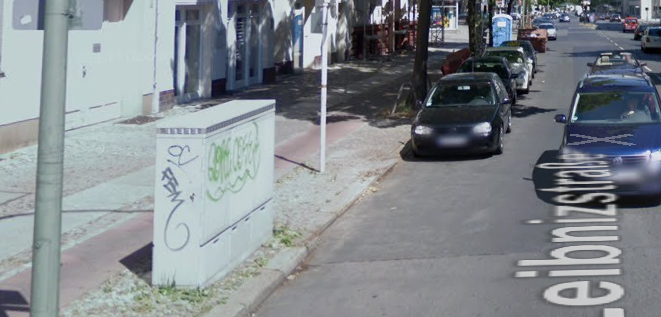}
	\caption{\leibniz is a low traffic neighborhood street with a separate bike lane which leads to relatively few near miss incidents (screenshot from Google Streetview).}
	\label{fig:leibniz}
\end{figure}

\textbf{\lei:}
Based on the collected data, \leibniz appears to be fairly safe.
We also analyzed the written incident descriptions in comparison to what we found on-site.
The street segment has a bike lane (though in poor condition) separated from street traffic, right next to the sidewalk (in the left part of figure~\ref{fig:leibniz}).
As such, it is not unexpected to find few incidents that are caused by having a shared bike/car infrastructure.
Nevertheless, we see the usual conflicts with pedestrians and the occasional left/right hook of a car driver.
Fortunately, the latter is uncommon as the other streets at intersections are low traffic neighborhood streets.
Overall, we conclude that this street segment is indeed relatively safe; safety in intersections and separation from pedestrians, however, could be improved.

\begin{figure}[t]
	\center
	\includegraphics[width=0.9\columnwidth]{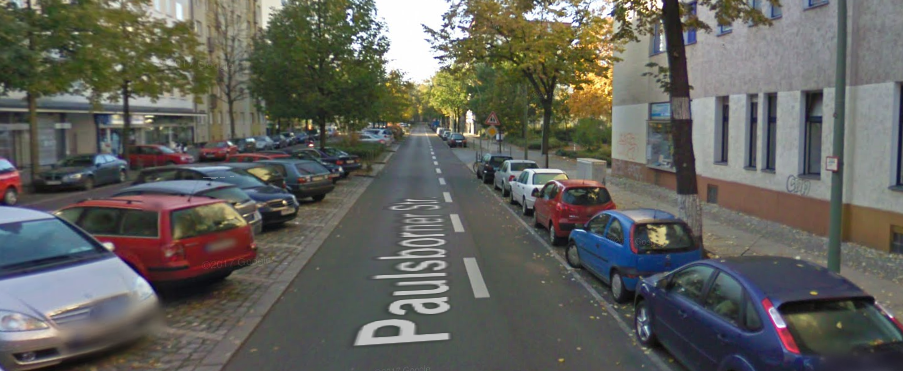}
	\caption{On \pau, cyclists and cars have to share the road which explains the high amount of close passes (screenshot from Google Streetview).}
	\label{fig:paul}
\end{figure}

\textbf{\pau:}
Based on the collected data, \paul appears to be a bit more dangerous than \leibniz but not in a way comparable to \edi.
The main problem, which we also identified based on the written descriptions, is close passes.
When looking at the Streetview Photo in figure~\ref{fig:paul}, the reason for this becomes obvious:
Cyclists traveling in the bike lane on the street tend to ride almost on the dashed line to avoid the dangerous dooring area next to the parked cars on the right.
This, however, means that any car passing the cyclist will keep a distance between 0.5m and 1m which qualifies as a close pass (and is also illegal in Germany).
Nevertheless, the dashed line suggests to drivers that it is safe to pass the cyclist.
While we did not measure, it appears to be impossible to pass a cyclist in a car (no matter where the cyclist is positioned) without violating the minimum passing distance as the car lane is rather narrow.

Contrary to the score metric, this street segment should therefore be classified as a dangerous place even though it is not an incident hotspot.
We believe that this is due to the fact that close passes cannot be detected by our measurement approach and need to be marked manually.
In conversation with users, we have heard that many users mark only critical close passes as there are too many of them in daily cycling.
This is also in line with the findings of Aldred and Goodman who discovered that 37\% of all observed incidents were close passes with 98\% being caused by motor vehicles and about half of them being perceived as ``on purpose''.
We believe that this is a limitation of our approach which we discuss in more detail in section~\ref{sec:disc}.
As such, we have to assume that our approach underestimates the number of close passes which we should consider in our detection of incident hotspots.
We plan to use the numbers of Aldred and Goodman to calculate an adjusted number of close passes.

%% file: content/5b_discussion.tex
Our approach has a number of inherent limitations which we will discuss in this section.

First, only a limited number of incident types can be detected based on the acceleration sensor measurements.
Specifically, our approach can only detect situations which caused sudden movement of the cyclist, e.g., swerving to the side or braking.
This implies that our approach can only detect close passes in case of really inexperienced cyclists who are not yet used to being threatened that way or when the close pass is a very narrow miss.
Likewise, tailgating is not detectable directly unless the smartphone is in a jacket pocket and the cyclist keeps turning around as this results in a repeated circular motion.
We counter this limitation by having the manual labeling and annotation phase after each ride which allows users to add such incidents.
These, however, will rarely be complete as there will be an (individual) upper limit on the number of incidents that a cyclist can remember which is especially problematic on long rides with many incidents.
As such, our data is very likely to underestimate the actual number of incidents.
We are currently considering options for a feature that offers live tagging, e.g., pressing a hardware button, or selecting the category directly on the smartphone screen during the ride.
While additional sensors would naturally improve the overall data quality, this inherently limits the number of participants.
Nevertheless, we are currently working on a feature that allows users to connect additional sensors to the smartphone app, e.g., distance sensors to detect close passes.

Second, one might argue that all incidents which we record are subjective and not objective.
This is indeed correct and desired as the \emph{perceived} safety, a subjective measure, is the key influence factor for mobility choices.

Third, our user group is not necessarily a representative subset of the cycling population.
In particular, we are unlikely to have senior citizens using our apps even though they may still be cycling.
We plan to compare our user group to official cycling statistics and derive corresponding aggregated incident metrics based on the profile data described in section~\ref{subsub:profile}.

Fourth, there is a margin of human error, due to the nature of crowdsourcing:
For instance, users may forget to stop the recording of their ride once the destination has been reached.
For that case, the mobile app's privacy slider can be used to crop the ride later on.
Even if the ride is uploaded without cropping, we can easily detect such a ride and crop it automatically.
Users could also record rides with a non-bicycle transportation mode, whether intentionally or not.
In most cases, we can also detect this easily based on the velocity of the ride -- there may, however, be false positives (e.g., with racing bikes) or false negatives (e.g., with electric scooters).
Finally, if incidents are mislabeled or not labeled at all, this can currently not be detected automatically.
We are currently exploring the use of machine learning methods for improved detection of incidents.
Overall, we believe SimRa can provide valuable insights.
While they may not be perfect, they are far better than anything available at the moment.

%% file: content/7_conclusion.tex
Increasing the modal share of bicycle traffic is a key mechanism to reduce traffic-related emissions but also to reduce traffic jams and the amount of space devoted to cars in inner city areas.
In practice, however, polls regularly show that a lack of (perceived) safety keeps people from using their bikes more frequently.
City planners who hence aim to improve bicycle safety need not only information on accidents but also on the much more frequent near miss incidents.
Such information, however, is not broadly available at the moment.

In this paper, we proposed to close that gap based on the crowdsourcing platform SimRa:
In SimRa, participants record their rides using a smartphone app which uses acceleration sensors to detect near miss incidents.
In a second step, users annotate and label the collected data before they upload it to our analytics backend.
There, we use exploratory data analysis methods based on a visualization toolkit as well as an automated scoring approach to identify incident hotspots.
Beyond the proposed platforms and the description of its deployment in Berlin since March 2019 (and other cities later), our contributions include the collected data set as well as first analysis insights.